\documentclass[twocolumn,aps,prl,preprintnumbers]{revtex4}
\usepackage[latin9]{inputenc}
\usepackage{amsmath}
\usepackage{amssymb}
\usepackage{graphicx}
\usepackage{mathrsfs}
\usepackage{amsfonts}
\usepackage{amsthm}
\usepackage{color,ulem}
\usepackage[colorlinks=true,citecolor=blue,linkcolor=blue,urlcolor=blue,anchorcolor=blue]{hyperref}%
\hypersetup{colorlinks=true,citecolor=blue,linkcolor=blue,urlcolor=blue}
\providecommand{\U}[1]{\protect\rule{.1in}{.1in}}
\makeatletter
\@ifundefined{textcolor}{}
{
\definecolor{BLACK}{gray}{0}
\definecolor{WHITE}{gray}{1}
\definecolor{RED}{rgb}{1,0,0}
\definecolor{GREEN}{rgb}{0,1,0}
\definecolor{BLUE}{rgb}{0,0,1}
\definecolor{CYAN}{cmyk}{1,0,0,0}
\definecolor{MAGENTA}{cmyk}{0,1,0,0}
\definecolor{YELLOW}{cmyk}{0,0,1,0}
}
\makeatother
\begin{document}
\title{Atomic antiferromagnetic domain wall propagation beyond the relativistic limit}
\author{Huanhuan Yang$^{1}$}
\author{H.Y. Yuan$^{2}$}
\email[]{yuanhy@sustc.edu.cn}
\author{Ming Yan$^{3}$}
\author{Peng Yan$^{1}$}
\email[]{yan@uestc.edu.cn}
\affiliation{$^{1}$School of Electronic Science and Engineering and State Key Laboratory of
Thin Films and Integrated Devices, University of Electronic Science and Technology of China,
Chengdu 610054, China}
\affiliation{$^{2}$Department of Physics, Southern University of Science
and Technology, Shenzhen 518055, Guangdong, China}
\affiliation{$^{3}$Department of Physics, Shanghai University, Shanghai 200444, China}
\date{\today}

\begin{abstract}
We theoretically investigate the dynamics of atomic domain walls (DWs) in
antiferromagnets driven by a spin-orbit field. For a DW with the width of a
few lattice constants, we identify a Peierls-like pinning effect, with the
depinning field exponentially decaying with the DW width, so that a spin-orbit
field moderately larger than the threshold can drive the propagation of an
atomic DW in a step-wise manner. For a broad DW, the Peierls pinning is negligibly
small. However, the external spin-orbit field can induce a fast DW propagation,
accompanied by a significant shrinking of its width down to atomic scales. Before
stepping into the pinning region, noticeable spin waves are emitted at the tail
of the DW. The spin-wave emission event not only broadens the effective width
of the DW, but also pushes the DW velocity over the magnonic barrier, which is
generally believed to be the relativistic limit of the DW speed.
 While the existing dynamic theory based on the continuum
approximation fails in the atomic scale, we develop an energy conversion theory
to interpret the DW dynamics beyond the relativistic limit.
\end{abstract}

\maketitle

{\it Introduction.---}Understanding the domain wall (DW) dynamics in magnetic
nanowires is of particular importance for both fundamental interest and application potential in spintronic devices \cite{Allwood2005,Parkin2008}.
In ferromagnets, the DW can be driven by an external field through the
dissipation of Zeeman energy \cite{Walker1974, Wang2009}, by an electric current
through the spin transfer torque (STT) \cite{SZhang2004, Tatara2004,Thiaville2005,Klaui2005}
and/or the spin-orbit torque (SOT) \cite{Manchon2009, Miron2011,Emori2013}, and by a spin wave
\cite{PYan2012,xiansi2012,Weiwei2015} and/or a thermal gradient through the magnonic angular momentum transfer \cite{Kovalev2012,Jiang2013,Nowak2014,Xiansi2014}.
The DW velocity typically increases with the external driving force at first and
then drastically decreases over a critical value, known as the Walker breakdown \cite{Walker1974,yuan2016},
because the transverse anisotropy fails to compete with the external force to sustain a stable planar DW propagation. Such a breakdown largely limits the maximum DW speed and therefore hinders the application
potential \cite{Beach2009}, although a cylindrical geometry \cite{MYan2010} was
recently proposed to overcome this issue, but yet to be realized experimentally.

Antiferromagnetic DW is superior to its ferromagnetic counterpart because of the absence of
the breakdown. Its propagation speed can keep increasing until the magnon velocity
close to tens of kilometers per second \cite{Helen2016,Shiino2016}. It has been shown that the DW width significantly shrinks when its velocity approaches the magnonic barrier, known as the Lorentz contraction effect \cite{Haldane1983,Yuan2018a, Kim2014}. The antiferromagnetic DW dynamics thus resembles the relativistic motion of a classical particle. However, the DW width should be limited by the intrinsic lattice constant due to
 the discrete nature of the crystal. Then it is intriguing to ask whether there
 is a generic cut-off of DW velocity in the antiferromagnet crystal. The static
 and dynamic properties of atomic DWs in ferromagnets have been well studied
 \cite{Hil1972,Novoselov2003,Yan2012}. However, the antiferromagnetic DW in
 such atomic scales receives little attention by the community.

In this work, we take the first step to theoretically study the behaviors of
atomic antiferromagnetic DWs under a spin-orbit field. For a static DW in atomic scales,
a generic pinning effect is identified in a clean system without
any disorders or defects, which originates from the Peierls-like potential.
The critical field for a DW propagation scales exponentially with the DW width.
For a dynamic case, we find that the DW velocity is not limited by such an
intrinsic pinning. Surprisingly, it could move even faster than the magnon velocity,
which is in sharp contrast to general belief. Detailed analysis uncovers the critical role
of strong spin-wave excitations accompanying the DW tail.

\begin{figure}
\centering
\includegraphics[width=\columnwidth]{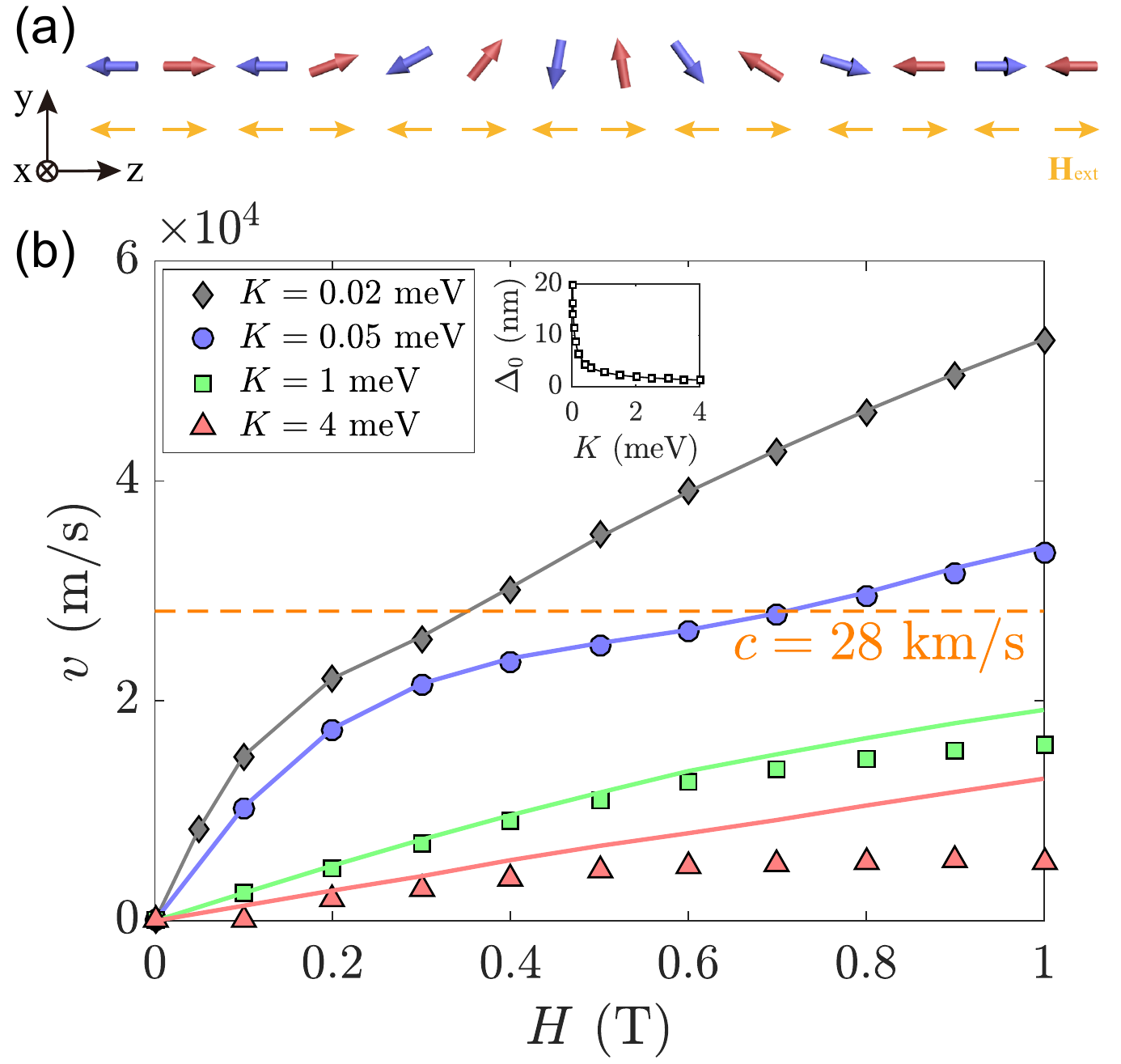}
\caption{(a) Schematic illustration of a two-sublattice antiferromagnet. The red and
blue arrows represent the magnetic moment on each sublattice,
respectively. Yellow arrows indicate the spin-orbit field. (b) The field dependence
of the DW velocity for $K = 0.02$ meV (gray rhombuses),
0.05 meV (blue dots), 1 meV (green squares), and 4 meV (red triangles).
The orange dashed line represents the relativistic limit.
(Inset) DW width as a function of the magnetic anisotropy. }
\label{fig1}
\end{figure}

{\it Model and main results.---}We consider a two-sublattice antiferromagnet
with easy-axis along the longitudinal direction ($z-$axis) as shown
in Fig. \ref{fig1}(a). A head-to-head DW locates at the center of the system
initially and then is driven to move under the action of a N\'{e}el field
($\mathbf{H}_a=-\mathbf{H}_b=He_z$) from the spin-orbit coupling \cite{Zelezny2014,Helen2016}.
Figure \ref{fig1}(b) shows the DW velocity as a function of field strength $H$
at various anisotropies while the inset shows the anisotropy dependence of the DW width.
Two important features can be identified: (i) When the anisotropy of the system is
stronger than $\sim4$ meV, there exists an intrinsic pinning field.
(ii) At smaller anisotropies, the DW velocity keeps increasing with field and goes
beyond the magnonic barrier (indicated by the dashed line), instead of being saturated,
in sharp contrast to common wisdom. Below, we examine these two features
in details.

{\it Intrinsic pinning.---}Let us first prove that the intrinsic pinning is
absent in a continuum theory, by starting from the Heisenberg Hamiltonian
\begin{equation}
\begin{aligned}
H_t=& J \sum_{\langle i,j \rangle} \mathbf{S}_{ai}\cdot
\mathbf{S}_{bj}-K \sum_i (\mathbf{S}_{ai,z}^2 +\mathbf{S}_{bi,z}^2) \\
&-\sum_i(\mathbf{S}_{ai}\cdot\mathbf{H}_a+\mathbf{S}_{bi} \cdot
\mathbf{H}_b),
\label{desh}
\end{aligned}
\end{equation}
where $\mathbf{S}_{ai}$ ($\mathbf{S}_{bj}$) ($|\mathbf{S}_{ai}|=|\mathbf{S}_{bj}|=S$)
are the spins on sublattices $a$ ($b$). $\langle i,j \rangle$ denotes the
nearest-neighboring sites. The first, second, and third terms in Eq. (\ref{desh}) are
the exchange coupling ($J>0$), magnetic anisotropy ($K >0$), and Zeeman energy,
respectively.

In terms of the magnetization $\mathbf{m}\equiv (\mathbf{S}_{ai}+
\mathbf{S}_{bi})/2S$ and the stagger order, $\mathbf{n}\equiv (\mathbf{S}_{ai}- \mathbf{S}_{bi})/2S$,
the Hamiltonian density $\mathcal{H}$ in continuum limit ($H_t\equiv\int (dz/d) \mathcal{H}$)
is given by \cite{Tveten2016, Yuan2018a},
\begin{equation}
\mathcal{H}= \frac{a}{2} \mathbf{m}^2 + \frac{A}{2} (\partial_z\mathbf{n})^2
-\frac{K_z}{2} n_z^2+ L  \mathbf{m} \cdot \partial_z \mathbf{n}
 - 2\mathbf{n} \cdot \mathbf{H},
 \label{conh}
\end{equation}
where $a\equiv 8JS^2$ and $A\equiv d^2 JS^2$ are, respectively, homogeneous
and inhomogeneous exchange constants with $d$ the lattice constant, $K_z\equiv 4KS^2$,
and $L \equiv 2dJS^2$ breaks the parity symmetry and results in a net magnetic moment
inside the AFM DW \cite{Pap1995,Yuan2018b}.

Minimizing the Hamiltonian with respect to the order parameters leads to the
following equation,
\begin{equation}
\frac{A}{2} \frac{\partial ^2 \theta}{\partial z^2}  - K_z \sin \theta \cos \theta -2H \sin \theta=0,
\label{static}
\end{equation}
where $\theta$ is the polar angle of the stagger order in spherical coordinates.
In the absence of external field, this equation naturally gives the Walker profile
\cite{Walker1974}: $\ln \tan (\theta/2)=z/\Delta_0$ with $\Delta_0=\sqrt{A/2K_z}$ the
static DW width. With external field, we can integrate both sides of the
equation and find that $H$ should be zero to validate Eq. (\ref{static}), which
 apparently contradicts with the finite field assumption. This implies that the
Eq. (\ref{static}) cannot be true under finite fields, i.e., a DW will
always move in a clean system without any disorders or notches.

\begin{figure}
\centering
\includegraphics[width=\columnwidth]{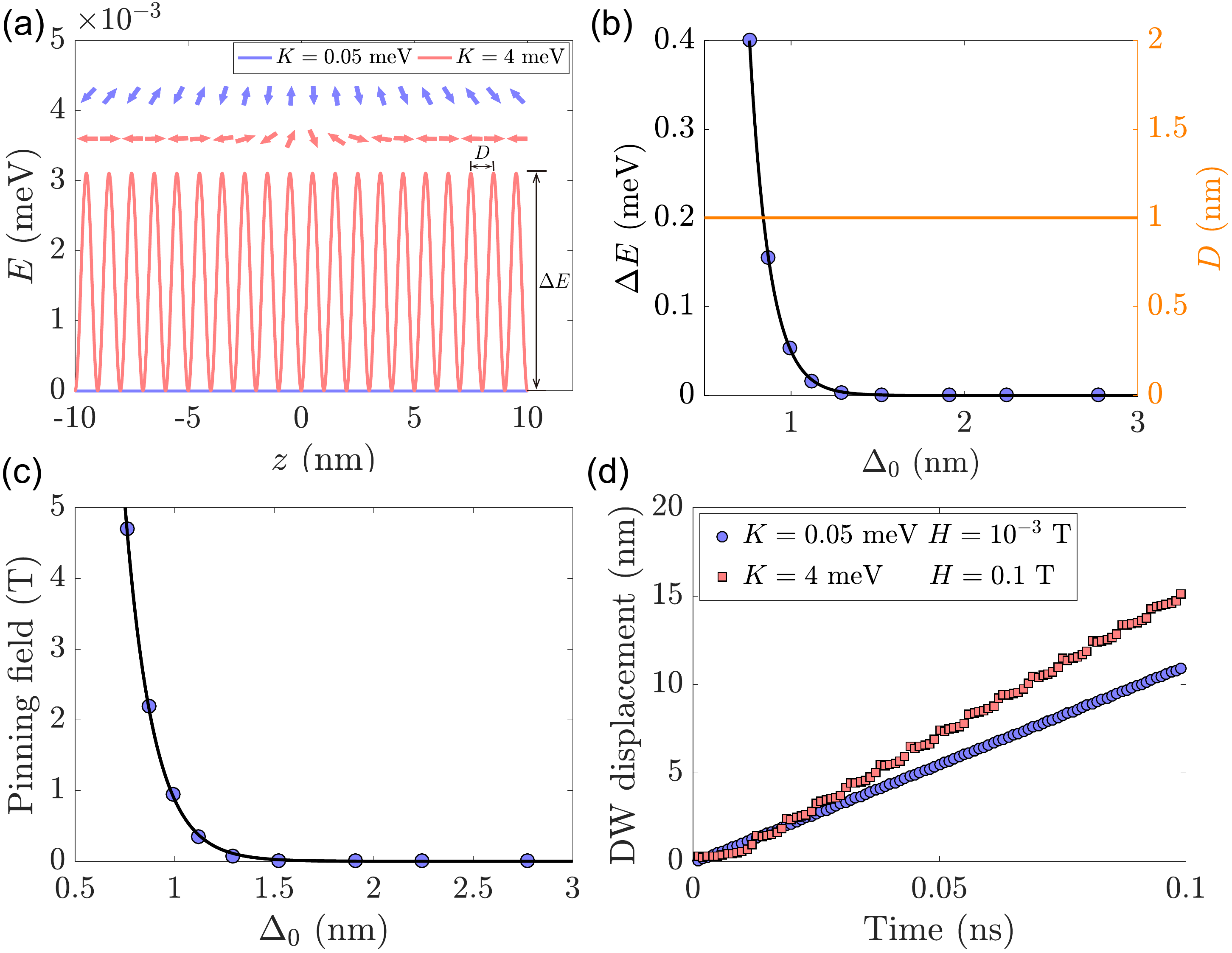}
\caption{(a) The Peierls-like energy profile of the DW when the DW center is
placed at different positions of the nanowire for $K = 0.05$ meV (blue line) and
$K = 4$ meV (red curve), respectively. (Inset) The corresponding spin profile of the static
domain wall. (b) The potential depth (blue dots) and period (orange line) as a function of DW width.
(c) The scaling of the pinning field as a function of DW width. The black curve is the theoretical fitting.
(d) The time dependence of DW displacement for $K = 0.05$ meV (blue dots) and $K = 4$ meV
(red squares), respectively. The DW position is defined as the position of spin with $n_y=1$.}
\label{fig2}
\end{figure}

Note that the absence of pinning is purely concluded from the continuum model, which is
well justified only if the length scale of the magnetic structure is much larger than the
lattice constant, i.e., $\Delta_0 \gg d$. This is not essentially true for atomic DWs.
To illustrate this point, we plot the energy of the system as a function of DW
position in Fig. \ref{fig2}(a), for various DW width. Clearly, as the DW width approaches
the lattice constant, the DW energy becomes strongly dependent on the DW position, which
provides a generic pinning potential for the DW motion, known as the magnetic Peierls potential \cite{Hil1972}.
The static profile of DWs should be the solution of $\delta H_t/\delta S_i=0$, i.e.,
\begin{equation}
\begin{aligned}
J(S_{b,i-1} + S_{b,i})-2K S_{ai}-H=0,\\
J(S_{a,i-1} + S_{a,i})-2K S_{bi}+H=0.
\end{aligned}
\label{disdw}
\end{equation}

The inset of Fig. \ref{fig2}(a) shows the spin configuration obtained by numerically
solving Eqs. (\ref{disdw}), where the DW center locates between two
nearest cells, i.e., the DW is trapped at the potential well in the
energy landscape. This explains why there exists an intrinsic pinning for a static DW.
Figure \ref{fig2}(b) shows that the pinning potential scales exponentially with
the DW width with a constant period $D$: $\Delta E=\Delta E_0 \exp[- \Delta_0/(\zeta D)]$,
where $\Delta E_0 =293$ meV and $\zeta=0.12$
are two fitting parameters. Figure \ref{fig2}(c) plots the DW-width dependence of
the depinning field, which can also be well described by the exponential function
$H_c = H_{c0} \exp[-\Delta_0/(\zeta D)]$ with coefficients $H_{c0}=986 ~\mathrm{T}$ and $\zeta=0.14$.

Naturally, it is expected that a DW should move in a step-wise manner in such a
periodic potential, which is indeed the case as shown in Fig. \ref{fig2}(d). Similar behavior
for ferromagnetic DWs has been theoretically predicted and experimentally verified
\cite{Hil1972,Novoselov2003}. For a wide DW, the step-wise trajectory
disappears.

\begin{figure}
\centering
\includegraphics[width=\columnwidth]{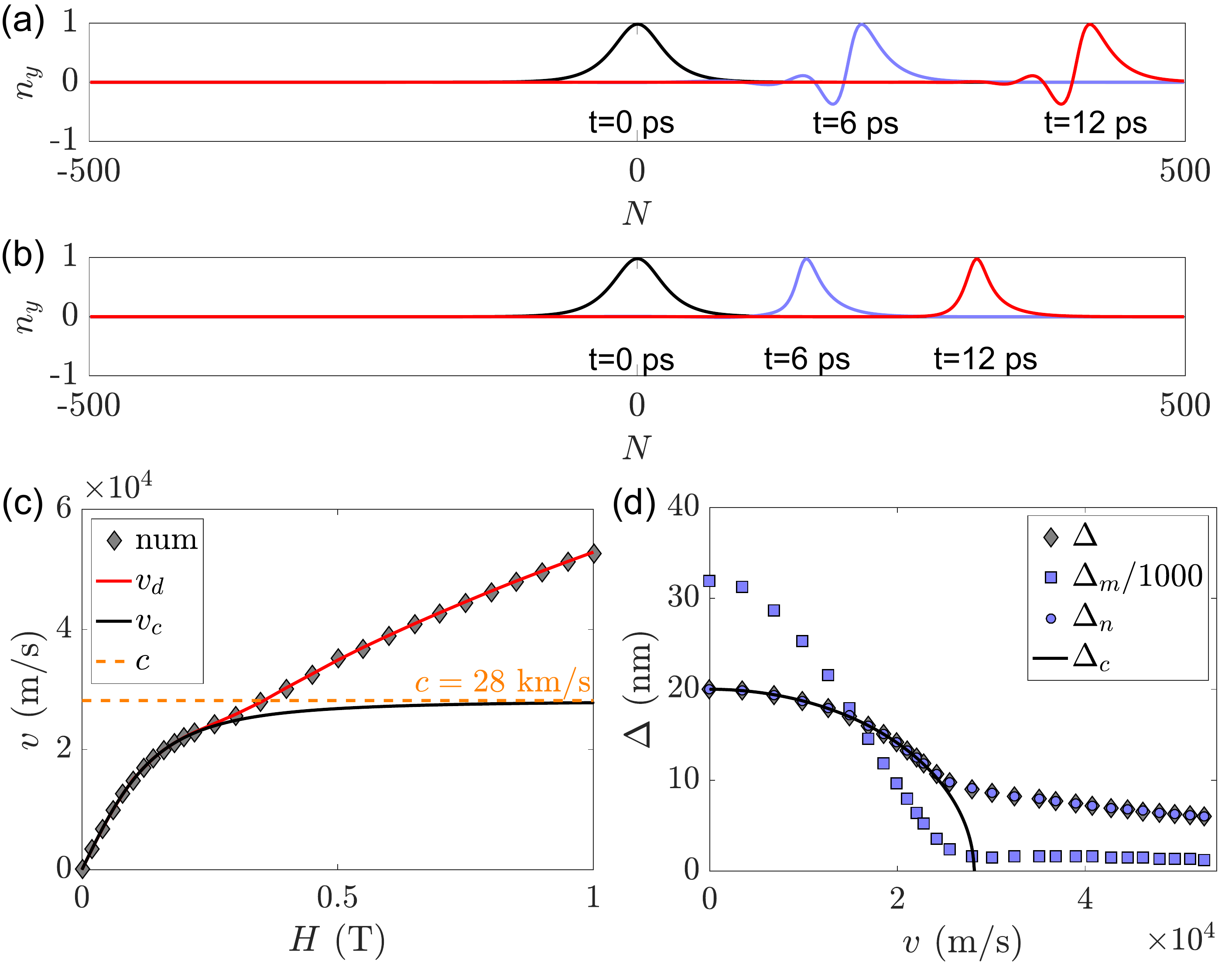}
\caption{(a) The profiles of magnetic structures at $t=0, 6, 12$ ps,
for $H= 0.5$ T (a) and 0.3 T (b), respectively.
(c) Domain wall velocity as a function of the field. Symbols are numerical results
and the red and black curves represent the prediction of a discrete model and continuous model,
respectively. The subscripts of $v_d$ and $v_c$ refer to the prediction of discrete
and continuous model, respectively. The orange dashed line is the magnon velocity
in this system. The parameters are $\Delta_0= 20$ nm and $\alpha=0.02$.
(d) The velocity dependence of the DW width.}
\label{fig3}
\end{figure}

{\it Beyond the magnonic barrier.}---Next we go to the dynamics of atomically narrow DWs.
The trivial case is an atomically narrow DW trapped by the
Peierls potential. However, the highly nontrivial case is that a wide DW could drastically
shrink its width to an atomic scale as its speed approaches the relativistic limit,
due to the Lorentz contraction. In what follows, we focus on the latter case.

As briefly introduced in Fig. \ref{fig1}(b), the DW could propagate beyond the
relativistic limit (or magnonic barrier), which suggests that the intrinsic pinning from the discrete
lattice does not play a significant role for dynamical DWs. To probe this propagation mode,
we plot the spin configurations at different times in Fig. \ref{fig3}(a),
with two important features: (i) The magnetic structure behaves as a soliton
or rigid-body that keeps its shape unchanged during the stable propagation.
(ii) The magnetization profile is no longer the Walker solution, where significant
spin-wave packets are emitted only at the tail of the wall. As a comparison,
Fig. \ref{fig3}(b) shows the case without spin-wave emission when the DW velocity
is well below the magnonic barrier. The DW dynamics above/below the relativistic
limit is reminiscent of the spin Cherenkov effect in a ferromagnetic cylinder \cite{MYan2011}.

Before presenting our theory on the DW motion beyond the magnonic limit, we first
discuss the analytics in the continuous limit, where the steady DW motion
obeys the sine-Gordon equation \cite{Yuan2018a}
\begin{equation}
\frac{1}{c^2}\frac{\partial^{2}\varphi}{\partial t^{2}} -\frac{\partial^{2}\varphi}{\partial z^{2}}
+ \frac{1}{\Delta_0^2}\sin \varphi =0,
\label{sinG}
\end{equation}
with $\varphi = 2\theta$ and $c=\gamma \sqrt{Aa/2}$ the magnon velocity. The 1-soliton
solution (DW) takes the general form $\varphi = 4\arctan \exp [(z-vt)/\Delta ]$
with $\Delta = \Delta_0 \sqrt{1-v^2/c^2}$. Then one can immediately see that
the maximum speed of the soliton is the magnon velocity when the
DW width approaches zero. Our results obviously cannot be explained by the sine-Gordon theory.
Tracing the numerical findings shown in Fig. \ref{fig3}(a), the spatial extension
of DW has reached atomic scales at high speeds, such that the spatial differential
operators in Eq. (\ref{conh}) and Eq. (\ref{sinG}) become not well
defined to replace the difference operators in Eq. (\ref{desh}).

It suggests us to abandon the continuous Hamiltonian (\ref{conh}) and the
sine-Gordon equation (\ref{sinG}) as well. The new starting point is the energy
conversion and its conservation. Regardless of the detailed DW profile, the change rate of
the Zeeman energy during the DW propagation should be equal to the
magnetic energy dissipation rate through the Gilbert damping, i.e.,
\begin{equation}
-2HM_s v= -\frac{\alpha M_s}{2\gamma} \sum_i \left [\left ( \frac{\partial \mathbf{S}_{ai}}{\partial t} \right )^2
+ \left ( \frac{\partial \mathbf{S}_{bi}}{\partial t} \right )^2 \right ],
\end{equation}
where the sum is taken over all atoms. For a rigid-body motion, we have $\partial_t \mathbf{S}_{\mu i}
 =-\mathbf{v} \cdot \nabla_l \mathbf{S}_{\mu i}$, where $\mu=a,b$, and $\nabla_l$
 is the first-order difference operator as $\nabla_l\mathbf{S}_{\mu i}
 =(\mathbf{S}_{\mu i}-\mathbf{S}_{\mu i-1})/d$. Then we can explicitly
express the DW velocity as
\begin{equation}
v=\frac{\gamma H \Delta_{\mathrm{eff}}}{\alpha }, \text{with}\ \Delta_{\mathrm{eff}}
= 4 \left [\sum_{\mu i} \left ( \nabla_l \mathbf{S}_{\mu i} \right )^2\right ]^{-1}.
\label{DWVelocityG}
\end{equation}
In the wide DW limit ($\Delta_{\mathrm{eff}} \gg d$),
we can replace the sum by the integral and the magnetic moments by the magnetization
order and the stagger order, and we obtain $\Delta^{-1}_{\mathrm{eff}}
= 2\int_\Omega [\left ( \partial_z \mathbf{m} \right )^2
+ \left ( \partial_z \mathbf{n} \right )^2]dz $. By disregarding the
quadratic term of the magnetization and adopting the Walker approximation,
we can show $\Delta_{\mathrm{eff}} = 2/\int_\Omega \left ( \partial_z \mathbf{n} \right )^2dz=\Delta $.
 Combining the Lorentz contraction $\Delta = \Delta_0\sqrt{1-v^2/c^2}$ with the
 velocity expression $v =\gamma H\Delta/\alpha$, we obtain the DW velocity as
$v=c\gamma H\Delta_0/\sqrt{\gamma^2 H^2 \Delta_0^2 + \alpha^2 c^2}$ which increases
as the spin-orbit field increases but finally saturates [as plotted by the black
curve in Fig. \ref{fig3}(c)].

When $\Delta_{\mathrm{eff}} \sim d$, the above velocity formula is not applicable,
but Eq. (\ref{DWVelocityG}) still holds. Figure \ref{fig3}(c) shows
that the theoretical formula (red curve) compares perfectly with numerical
results over the entire spin-orbit field regime. To clarify the variation
of whether the magnetization or the stagger
order dominates the DW velocity in high fields, we define
two DW width $\Delta_n$ and $\Delta_m$ in terms of $\Delta_m=2\left [ \sum_i \left (\nabla_l\mathbf{m}_{i} \right )^2
\right ]^{-1},\Delta_n =2\left [ \sum_i \left (\nabla_l\mathbf{n}_{i} \right )^2 \right ]^{-1}$.
Then the DW velocity can be rewritten as

\begin{equation}
v=\frac{\gamma H }{\alpha} \left (\frac{1}{\Delta_m}+ \frac{1}{\Delta_n} \right )^{-1}.
\end{equation}

Figure \ref{fig3}(d) shows that $\Delta_m$ is almost
three orders of magnitude larger than $\Delta_n$, and its contribution to the DW propagation
can therefore be neglected. Moreover, $\Delta_n$ in the high velocity
regime approaches a constant value around 5 nm instead of zero as predicted by the
continuum field theory. This difference comes from the spin-wave excitation at
the tail of an atomic DW, which effectively broadens the DW width.

\begin{figure}
\centering
\includegraphics[width=\columnwidth]{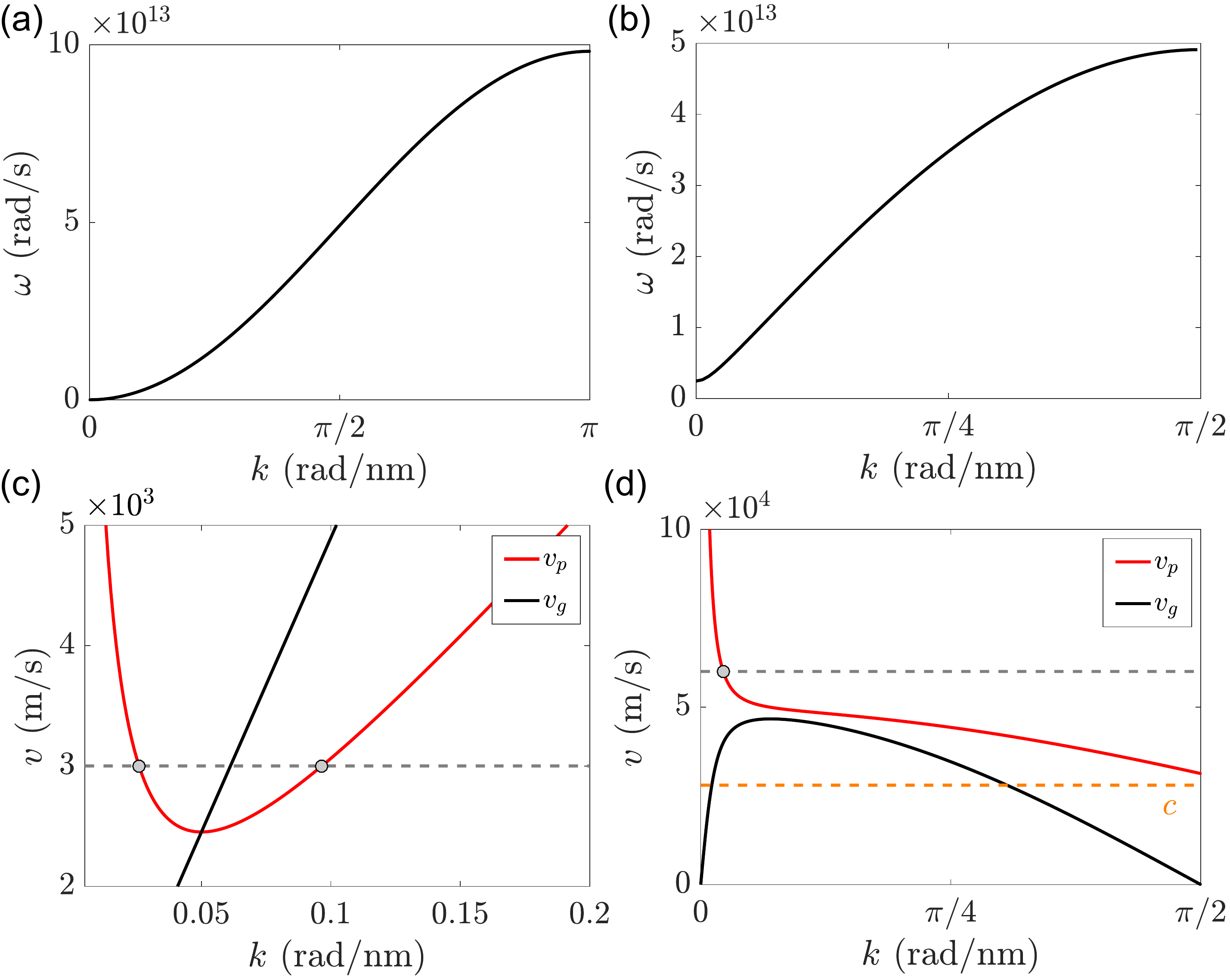}
\caption{Spin-wave spectrum in ferromagnets (a) and antiferromagnets
(b). $k-$dependence of the phase velocity (red curve) and the group velocity (black curve)
for ferromagnetic (c) and antiferromagnetic spin-waves (d),
respectively. Parameters $J=16$ meV and $K=0.02$ meV are used for both
ferromagnets and antiferromagnets, for a fair comparison. The black (orange) dashed
lines refers to a speed beyond (at) the relativistic limit \cite{note01}. }
\label{fig4}
\end{figure}

{\it Discussions and Conclusions.---}Firstly, we would like to compare our results
with the supermagnonic observation
in a cylindrical ferromagnetic nanowire, where spin waves are emitted both in the
front and tail of the DWs with different frequency but share an equal phase
velocity \cite{MYan2011}. In our case, spin waves are only emitted at the tail of the DW.
This difference can be understood from the distinct nature of ferromagnetic and
antiferromagnetic spin-wave dispersions, as shown in Figs. \ref{fig4}(a) and (b).
In a ferromagnet, there exist two spin-wave modes for a given phase speed (DW speed) while
the group velocity of the high (low) frequency mode is larger (smaller) than the
DW speed [as plotted in Fig. \ref{fig4}(c)]. Hence the high-frequency mode runs in front of the
DW while the low-frequency one lags behind. For an antiferromagnet, only
one spin-wave mode exists at a given phase speed and its group velocity is
always smaller than its phase velocity (DW speed) [see Fig. \ref{fig4}(d)]. So it is always lagged
behind the wall. Further, from the perspective of materials, a cylindrical geometry
is required to eliminate the Walker breakdown effect and to accelerate the DW into
the magnonic regime, while it is still a challenge to make perfect cylindrical magnetic
wire and to detect the nanoscale DW motions in particular.
However, for antiferromagnets, a planar film is sufficient to reach this magnonic regime
with a typical DW velocity as high as several tens of km/s, 10 times faster than the speed in
ferromagnets.

Secondly, the dissipation mechanism of spins is a long-lasting issue in general magnetism.
Recent first-principles calculations on metallic antiferromagnets shows that
the damping through the magnetization order ($\alpha_m$) could be 2-3 orders of magnitude
stronger than that through the stagger order ($\alpha_n$) \cite{Yuan20171,Mah2018}.
One of the present authors contributes to this outstanding finding by uncovering the role of
inter-sublattice spin pumping effect $\alpha_c$ \cite{Yuan20171,Yuan20172}, where
$\alpha_m = \alpha + \alpha_c$ and $\alpha_n=\alpha - \alpha_c$. Following the energy conversion and conservation
approach, we envision the general form of DW velocity as $v=\gamma H \left (\alpha_m/\Delta_m+ \alpha_n/\Delta_n \right )^{-1}$.
Then the term $\alpha_m/\Delta_m$ could be comparable with $\alpha_n/\Delta_n$,
if $\alpha_c\sim \alpha$. Our result thus suggests a new route to determine the damping
parameters by measuring the atomic DW propagation velocity in experiments.

In conclusion, we have investigated both the static and the dynamic behaviors of atomic DWs in
antiferromagnets. An intrinsic pinning effect for the atomic DW was identified, due to the
discrete nature of crystal lattices instead of disorders or defects. A dynamic subatomic DW is
influenced very little by this pinning effect and its velocity
could go beyond the relativistic limit. An energy conversion and conservation theory beyond the
conventional sine-Gordon approach was developed to explain the numerical findings.
Our results provide a new way to experimentally resolve the magnetization damping
and stagger order damping in antiferromagnets.

\acknowledgments
{\it Acknowledgments.---}
HYY is financially supported by National Natural Science Foundation of China
(NSFC) Grants (No. 61704071).  PY was partially supported by NSFC Grant
(No. 11604041). MY is supported by NSFC Grant (No. 11774218).

\end{document}